\documentclass[a4paper,12pt]{spieman}  
\usepackage{amsmath,amsfonts,amssymb}
\usepackage{graphicx}
\usepackage{setspace}
\usepackage{tocloft}
\usepackage{graphicx}	
\usepackage[a4paper, total={6.75in, 8.5in}]{geometry}

\newcommand{\pvol}{P$_{vol}$ }
\newcommand{\ptr}{P$_{tr}$ }
\newcommand{\pfov}{P$_{fov}$ }
\newcommand{\sqdeg}{deg$^2$ }

\title{GERry: A Code to Optimise the Hunt for the Electromagnetic Counter-parts to Gravitational Wave Events.}

\author[a]{David O'Neill}
\author[a]{Joseph Lyman}
\author[a]{Kendall Ackley}
\author[a]{Danny Steeghs}
\author[b]{Duncan Galloway}
\author[c,d]{Vik Dhillon}
\author[e]{Paul O'Brien}
\author[f]{Gavin Ramsay}
\author[g]{Kanthanakorn Noysena}
\author[h]{Rubina Kotak}
\author[i]{Rene Breton}
\author[j]{Laura Nuttall}
\author[d]{Enric Pall\'e}
\author[a]{Don Pollacco}
\author[a]{Krzysztof Ulaczyk}
\author[c]{Martin Dyer}
\author[b]{Felipe Jim\'enez-Ibarra}
\author[g]{Tom Killestein}
\author[a]{Amit Kumar}
\author[j]{Lisa Kelsey}
\author[a]{Ben Godson}
\author[c]{Dan Jarvis}
\author[ ]{and the GOTO Collaboration.}

\affil[a]{Department of Physics, University of Warwick, Gibbet Hill Road, Coventry CV4 7AL, UK}
\affil[b]{School of Physics \& Astronomy, Monash University, Clayton VIC 3800, Australia}
\affil[c]{Department of Physics and Astronomy, University of Sheffield, Sheffield S3 7RH, UK}
\affil[d]{Instituto de Astrof \'isica de Canarias, E-38205 La Laguna, Tenerife, Spain}
\affil[e]{School of Physics \& Astronomy, University of Leicester, University Road, Leicester LE1 7RH, UK}
\affil[f]{Armagh Observatory \& Planetarium, College Hill, Armagh, BT61 9DG, UK}
\affil[g]{National Astronomical Research Institute of Thailand, 260 Moo 4, T. Donkaew, A. Maerim, Chiangmai, 50180 Thailand}
\affil[h]{Department of Physics \& Astronomy, University of Turku, Vesilinnantie 5, Turku, FI-20014, Finland}
\affil[i]{Jodrell Bank Centre for Astrophysics, Department of Physics and Astronomy, The University of Manchester, Manchester M13 9PL, UK}
\affil[j]{Institute of Cosmology \& Gravitation, University of Portsmouth, Portsmouth PO1 3FX, UK}

\cftpagenumbersoff{figure}
\cftpagenumbersoff{table} 
\begin{document} 
\maketitle

\begin{abstract}
The search for the electromagnetic counterparts to gravitational wave (GW) events has been rapidly gathering pace in recent years thanks to the increasing number and capabilities of both gravitational wave detectors and wide field survey telescopes. Difficulties remain, however, in detecting these counterparts due to their inherent scarcity, faintness and rapidly evolving nature. To find these counterparts, it is important that one optimises the observing strategy for their recovery. This can be difficult due to the large number of potential variables at play. Such follow-up campaigns are also capable of detecting hundreds or potentially thousands of unrelated transients, particularly for GW events with poor localisation. Even if the observations are capable of detecting a counterpart, finding it among the numerous contaminants can prove challenging. Here we present the Gravitational wave Electromagnetic RecoveRY code (\textsc{GERry}) to perform detailed analysis and survey-agnostic quantification of observing campaigns attempting to recover electromagnetic counterparts. \textsc{GERry} considers the campaign's spatial, temporal and wavelength coverage, in addition to Galactic extinction and the expected counterpart light curve evolution from the GW 3D localisation volume. It returns quantified statistics that can be used to: determine the probability of having detected the counterpart, identify the most promising sources, and assess and refine strategy. Here we demonstrate the code to look at the performance and parameter space probed by current and upcoming wide-field surveys such as GOTO \& VRO.
\end{abstract}

\keywords{Multi-messenger astronomy,strategy,survey}

{\noindent \footnotesize\textbf{*}David O'Neill,  \linkable{David.O-Neill@Warwick.ac.uk} }

\begin{spacing}{1}   

\section{Introduction}
\label{sect:intro}  

Astrophysics has long relied almost solely on the electro-magnetic force to inform us about the Universe. The first detection of gravitational waves (GW) by LIGO in 2015 however, paved the way for a new era of routine multi-messenger astrophysics through the combination of traditional electromagnetic (EM) datasets complemented by GW information. The current array of GW inferometers are capable of detecting the mergers of stellar mass compact objects. While a combined EM and GW detection of binary black hole (BBH) mergers are not expected due to the expected lack of EM emission from such an event, neutron star - black hole (NSBH) and, in particular, binary neutron star (BNS) mergers are capable of producing an EM signal. Indeed, short gamma-ray bursts (SGRB) originate from the powerful relativistic outflows produced in BNS\cite{Abbot2017} and possibly from NSBH merger events\cite{BenNSBH}.

While SGRBs are regularly detected, they are not the only EM emission released from BNS mergers. The decay of r-process elements in the merger ejecta can also produce longer-lasting but faint kilonovae (KNe). However, the number of detected KNe remains very small, with only a handful of photometric candidates, and two spectroscopically confirmed cases: the notable KN AT~2017gfo (see the review by Margutti \& Chornock\cite{KN17gfo}) and a KN associated with the GRB 230307A\cite{LevanKN}.

This small sample is not only driven by the inherent low rates of BNS mergers, but also due to the faint and rapidly evolving nature of KNe which may only remain detectable in the optical for $\leq$5 days\cite{mosfit170817,KNdiversity}, compared with supernovae (SNe), which generally evolve over timescales of weeks to months.

The current array of survey telescopes such as ATLAS\cite{ATLAS}, BlackGEM\cite{BlackGEM}, DES\cite{DEcam}, GOTO\cite{GOTODanny,GOTOMartin}, PanStarrs\cite{panstarrs}, and ZTF\cite{ZTFBellm,ZTFGraham} typically have single or multiple mounts with a combined field of view on the order of 10--100s\,\sqdeg and provide high cadence data across much of the sky, whereas the upcoming 8.4m Vera C. Rubin Observatory\cite{VRO} sacrifices field of view and cadence ($\sim$ 9.6\,\sqdeg) for a much deeper imaging. Despite this current array of wide field instruments, KNe remain elusive due to their rapid, faint and rare nature. 


The distance and sky-localisation of GW events are sensitive to the number of detectors online during the event. With still a low number of detectors, GW localisations are typically heterogeneous, and place challenging requirements on EM follow-up campaigns, which often need to not only cover large sky areas, but do so at high cadence and with good sensitivity. Due to the variety of GW localisation possible, coupled with different EM signatures expected, identifying a single strategy to optimise recovery probability EM counterparts is complex.


In this paper, I show how the Gravitational wave Electromagnetic RecoveRY code (\textsc{GERry}) can help in these respects, by providing a method to rapidly provide an in depth statistical analysis of GW follow-up observations and returning the probability of detecting a KNe event that folds in observing footprints, temporal and wavelength coverage, Galactic extinction, and observing sensitivity in comparison to the expected counterpart temporal evolution. This can be used to determine whether a KN is likely to be among the detected sources as well as to test the effectiveness of the observing strategy for a given event.
Typically such analysis is performed long after any KN would have faded. By providing this information rapidly while the KN is still potentially visible, it can allow users to adapt their strategies and determine whether the mobilisation of further follow-up resources is required.

\section{Methods}

\subsection{Multi-Order Coverage skymaps}
One of the most popular methods of handling sky projections in modern astronomical data analysis is the use of Hierarchical Equal Area isoLatitude Pixelization (HEALPix\cite{healpix}). HEALPix allows the sky to be subdivided into pixels or tiles, the number of which is dependent on the desired resolution. This data format was adopted by LIGO for producing the probabilistic skymaps output by 3D localisation codes such as BAYESTAR and LALinference. While the HEALPix grid itself is 2D, the localisation codes encode the probability contained within each tile as well as the estimated distance in Mpc, thus turning the 2D HEALpix skymap into a 3D map.

This format was further built on by utilising an multi-resolution grid in the form of Multi-Order Coverage skymaps (MOCs\cite{mocs}). Originally developed to enable the rapid spatial/coverage queries such as intersections and unions on a HEALPix grid, it was incorporated into LIGO GW maps where credible regions/contours with higher probability densities are assigned higher HEALPix resolutions with the inverse being true for lower probability dense areas. The primary advantage of this format over the previous `flat' resolution maps is they are less computationally intensive to both handle and analyse.

\subsection{healpix-alchemy}
\textsc{GERry} is primarily built on \textsc{healpix-alchemy}\cite{Singer2022}, a Python package that enables rapid querying of GW MOCs using SQLAlchemy in conjunction with a PostgreSQL 14 database. \textsc{healpix-alchemy} utilises the multi-range data type in the form of a range set in PostgreSQL 14, whereby each HEALPix tile at a given resolution is represented by a \textit{rangeset} corresponding to the indices of the much smaller constituent tiles (sub-pixels) at the highest resolution at which the tile indices can be stored as 64bit integers. Given the sequential nature of these ``sub-pixels", one HEALPix tile at any resolution can always be represented by a single range set containing the minimum and maximum indices of the sub-pixels, which will take the form, e.g., [576460752303423488, 577586652210266112).  Therefore, the GW skymap can be broken down into an array of rangesets for each tile. These tiles are then stored in the database along with the tiles' probability density values, distance estimate and distance uncertainty. The same process is performed for the all image in the follow-up campaign, whereby the footprint of each image is mapped onto a multi-resolution HEALPix grid. The code employ MOC resolution of the constituent HEALPix tiles such that higher resolutions (smaller tiles) are at the edges and corners of the images and lower resolution tile (larger tiles) in the centre, in order to maintain the original image field of view as much as possible whilst reducing computational cost later.  
Now that both the probabilistic GW skymap and the images have been broken down into HEALPix tiles and stored in the database, their HEALPix rangesets can now be efficiently queried against each other. Given the sequential nature of the rangesets, if the range of one rangeset overlaps that of another then the tiles they represent also must overlap in sky position by an amount corresponding to this rangeset overlap. This effectively allows one to trivially find which image tiles overlap or contain GW skymap tiles, the size of the overlap, and the area and the probability contained within the image tiles (see Singer et al. 2022\cite{Singer2022} their figure 4).

\subsection{\textsc{GERry}: from 2D to 3D}
While \textsc{healpix-alchemy} allows the user to easily find the 2D properties of your follow-up campaign, \textsc{GERry} extends this to 3D by utilising the encoded distance information contained within the skymaps in combination with the sensitivity of the field images.
\textsc{GERry} operates under a simple concept, if the distance to the GW event at a particular sky position is known from the skymap, one can use this alongside the expected temporal evolution of the counterpart to estimate the expected apparent magnitude at the time any image of this sky position was taken. By converting the encoded distance distribution to a brightness distribution, the image depths then provide a means to calculate the fraction of this brightness distribution that was recovered, and thus the fraction of 3D probability recovered in each tile. These recovery fractions provide quantified values such as the total probability volume recovered (P$_{vol}$) or the probability of detecting the counterpart in at least once across the imaged sky (P$_{transient}$ or P$_{tr}$).
In the following subsections we show in detail how these values are calculated.

\subsection{General processing}

\subsubsection{Find image coverage and overlap}
The first step is to identify skymap tiles that are overlapped or contained wholly within an image by querying their respective rangesets for overlaps. Each skymap tile and overlapping image tile combination is stored in a \textsc{pandas}\footnote{https://pandas.pydata.org/pandas-docs/stable/index.html} dataframe alongside information that will be used in processing: skymap probability density, area, and distance information, and the image tile observing time, filter, and depth (and uncertainty).
Most skymap tiles are not contained entirely within a single image tile, therefore the area of the skymap tile covered by every partially overlapping image tile has to be calculated. For each skymap tile and overlapping-image tile combination, the overlap area is calculated using the size of the overlapping rangeset multiplied by the area of a single sub-pixel ($\sim$1.2$\times$10$^{-14}$ \sqdeg). The probability recovered by a particular skymap-image tile overlap is then the probability density multiplied by the overlap area.

\subsubsection{Generate counterpart magnitudes}

The total recovered 3D probability can not be known unless we estimate the light curve evolution of the counterpart. The probabilities of finding a KN-like counterpart will be a strong function of time since the event, due to their rapid evolution. The potential diversity of GW EM counterparts has yet to be mapped given that for KNe, GW170817-KN2017gfo remains the only KN event with extensive light curve coverage across the EM spectrum.
Until such a time we have a robust sample of KNe light curves, we rely on models. An input light curve file to \textsc{GERry} contains the phase and expected absolute magnitude and magnitude uncertainty of the counterpart. A separate light curve is required for all the image filters. \textsc{GERry} will calculate the counterpart phase at the time the image was taken (with respect to the GW event) and find the closest point in time in the given model light curve. Given that we are using model light curves, the density of photometric points can be arbitrarily increased in to improve temporal resolution. If an observed light curve is being used, we recommend interpolating between the points down to a time step on the order of $\sim$1--2 hours for rapidly evolving transients.

The apparent magnitude of the counterpart in a given skymap tile at a given time can then be found using the absolute magnitude and the embedded GW skymap distance and distance uncertainty information. Galactic foreground extinction is also calculated at the position of the skymap tile, (specifically the central coordinates of the tile) using the using the Planck collaboration dust map queried through the \textsc{dustmaps} Python package\cite{dustmaps} and applied to the estimated magnitude if the filter bandpass has been provided, this can also be turned off if desired. This process generates the expected magnitudes for the counterpart in all skymap-image tile overlaps.

\subsection{Polygons and Probabilities}
The counterpart recovery fraction at a particular sky position is calculated by modelling the counterpart's probability distributed over magnitude space as a normal cumulative distribution function with mean and standard deviation determined by the estimated magnitude and uncertainty. The total probability recovered at this position of the sky, $P_{int}$, is then the value of the cumulative distribution at the achieved image depth. The uncertainties are determined by repeating this process for the upper and lower image depth uncertainty bounds. This adds to the dataframe of skymap-image tile overlaps the recovered probability values. 

It is likely, however, that a skymap tile has been imaged multiple times, and it is imperative for the following calculations to know exactly what part of the skymap tiles each overlapping image tile physically covered. For example, two image tiles may overlap 50\% of a particular skymap tile -- this could mean 100\% or 50\% coverage of that skymap tile, depending how the image tiles are arranged on-sky, which naturally would affect recovery statistics.
In order to facilitate this, all overlapping image HEALPix rangesets for a particular skymap tile are merged and split into unique non-overlapping rangesets or "polygons" for each overlapping image combination.
This results in a list of unique skymap-image tile combinations to calculate the probability statistics such as \pvol and \ptr.

\pvol is the maximum 3D volume probed over the course of the campaign. For skymap tiles observed multiple times, only the most constraining observation (i.e the observation which recovered the most probability) is considered. It is the sum of the maximum probability achieved at each part of the imaged sky. A \pvol value of 100\% indicates that all of 3D probability space for a given event (assuming a given set of light curve models) has been recovered. In essence it is the often-quoted \pfov -- the sum of the 2D probabilities contained within the bounds of the images -- extended to 3D space. Unlike \pfov however, \pvol more informatively folds in the limitations of the images' sensitivities in the face of an expected counterpart's evolution.

\ptr is the probability of detecting the desired counterpart \textit{at least once} across the imaged sky. In cases where a region of sky has been imaged multiple times, \pvol only includes the probability contribution of only the most constraining tile. \ptr on the other hand combines the probability contribution from all overlapping image tiles. E.g. for a single skymap polygon $i$, the \ptr value would be: 
\begin{equation} 
\label{eq1}
P_{tr,poly_i} = P_{tot,poly_i}(1-[(1-P_{int_1,img_1})\times(1-P_{int_2,img_2})~...~\times(1-P_{int_n,img_n})])
\end{equation}

Where $P_{tr,poly}$ is the probability of detecting the counterpart at least once within this polygon, $P_{tot,poly}$ is the total 2D probability contained within the polygon, (the skymap probability density multiplied by the area of the polygon), and $P_{int_n,img_n}$ is the recovered probability of this polygon from the $n^{th}$ image tile. The final \ptr value is then simply the sum of all $P_{tr,poly}$ values for the total number of polygons $N$:
\begin{equation} 
P_{tr} = \sum^N P_{tr, poly_i}
\end{equation}
In cases where no images overlap in the follow-up campaign (and assuming a non-infinite image depth):
\begin{equation} 
P_{fov} \ge P_{tr} = P_{vol}
\end{equation}

For follow-up with overlapping images, the probability volume is being sampled more than once, therefore:
\begin{equation} 
P_{fov} \ge P_{tr} \ge P_{vol}
\end{equation}


Another output of \textsc{GERry} is the \ptr skymap. This skymap contains \ptr density in place of the standard localisation probability density. Since \ptr is the probability of detecting the EM counterpart at any position of the sky, it can be used for a much more focused search for EM counterpart candidates. The \ptr density will naturally be negligible for areas of the sky that suffer from e.g high dust extinction, poor observing conditions, imaging `too-late' after the event. By cross-matching detected sources with the \ptr skymap in place of the standard skymap, the search area can be properly prioritised for human-inspection of candidates, allowing for a much more rapid identification of any counterpart candidate.

\section{Detailed Example: S190425z}

Of LIGO's O3 observing run, S190425z was the only statistically significant BNS merger. At an estimated distance of $\sim$150 Mpc, it had the potential to produce a detectable, albeit faint, EM counterpart. However, localisation was poor ($\sim$9000 deg$^2$) due to only being detected in the Livingston (L1) and Virgo (V1) detectors, and much of the probability was spread across both northern and southern hemispheres. 

\begin{figure}
 \centering
\includegraphics[width=0.85\columnwidth]{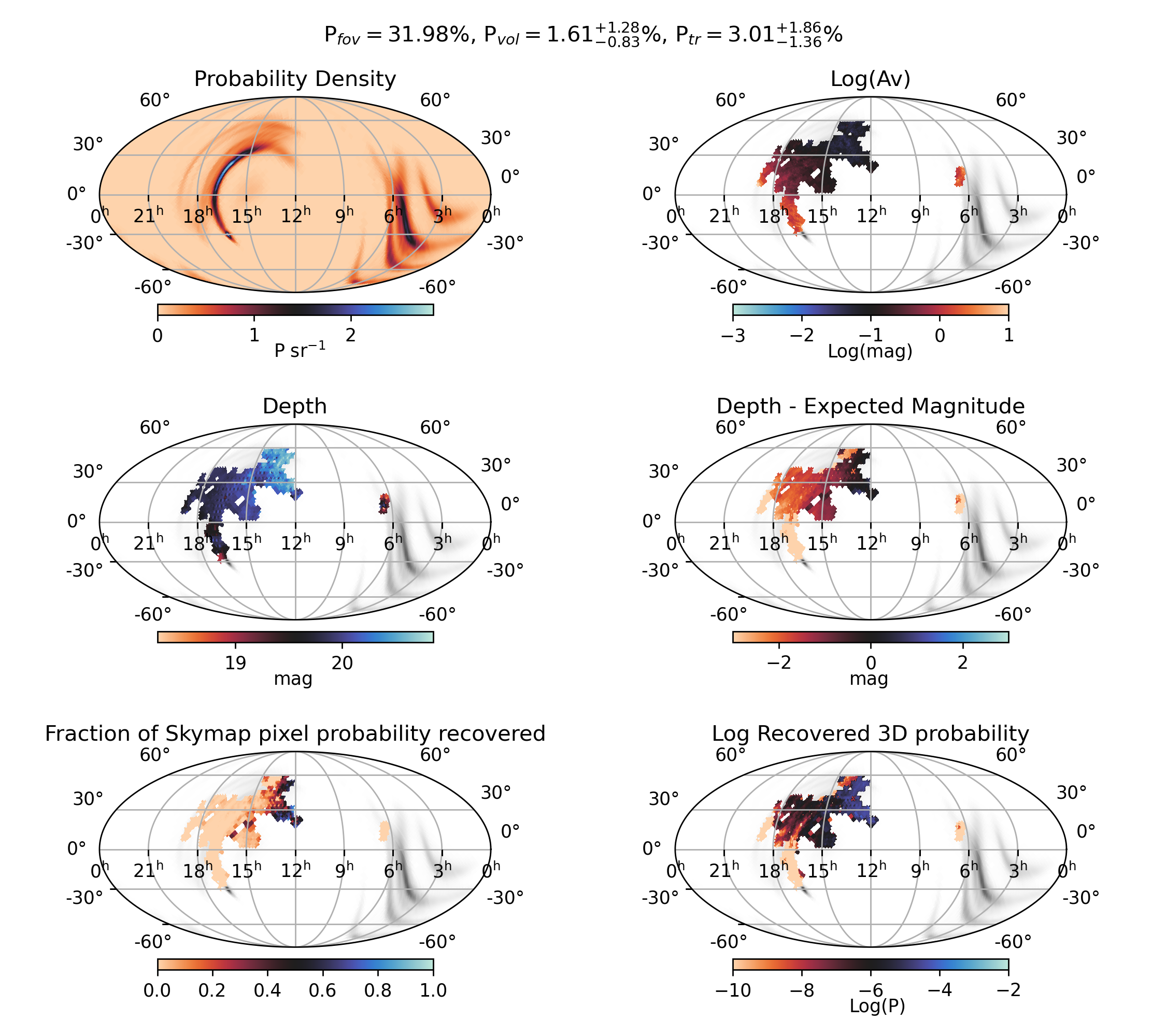}
    \caption{Skymap coverage information output by \textsc{GERry}. \textbf{Top left:} The standard skymap probability density information. \textbf{Top right:} Line of sight extinction of all imaged areas of the sky. \textbf{Mid left:} The average depth achieved at each part of the imaged sky. \textbf{Mid right}: The difference between the depth and expected counterpart magnitude. \textbf{Bottom left:} The ratio of the recovered 3D tile probability P$_{int,tile}$ to the total probability contained within the tile P$_{tot,tile}$. \textbf{Bottom right:} The maximum 3D probability recovered in each tile.}
    \label{fig:S190425z_skymap_plots}
\end{figure}

At the time, a GOTO prototype system (GOTO-4) was in operation\cite{GOTODanny,GOTOproto}, it consisted of a single mount at Roque de los Muchachos observatory in La Palma, and was equipped with four 0.4m telescopes with a combined field-of-view of $\sim$20~\sqdeg.
GOTO observations of the skymap spanned approximately 3 days from 2019-04-25 20:40 to 2019-04-28 05:15 (+0.5 to 2.8 days post trigger) and consists of 1948 L-band (400--700~nm) images covering 2155~\sqdeg. Figure \ref{fig:S190425z_skymap_plots} shows the outputs of \textsc{GERry} of this observing campaign, including the skymap, observed regions, and spatially resolved reddening, image depth, and recovered probability information. The total 2D probability covered was \pfov= 32\% owing to much of the probability being in-accessible to the GOTO-4 prototype. The light curve used was generated with \textsc{MOSFiT}\cite{mosfit} a python package typically used to fit semi-analytical light curves to photometry points for parameter estimation. Here, we use it in the 'generative' mode, where MOSFiT draws samples from a given list of parameter priors and generates model light curves in the desired filters. For S190425z, we use the input parameters detailed in Smartt et al.~(2023)\cite{Smartt190425}.
Like AT~2017gfo, the expected counterpart similarly peaks and fades rapidly within 24 hours, albeit $\sim$0.7-1.0 mag fainter.
Using this light curve we find the total probability volume probed by the GOTO prototype was \pvol=$1.61^{+1.28}_{-0.83}$\% with \ptr=$3.01^{+1.86}_{-1.36}$\%, the coverage and depth is comparable to that achieved by other surveys and serves to highlight the challenges of such follow-up, particularly when large areas of the skymap are inaccessible.

\begin{figure}
\centering
\includegraphics[width=0.75\columnwidth]{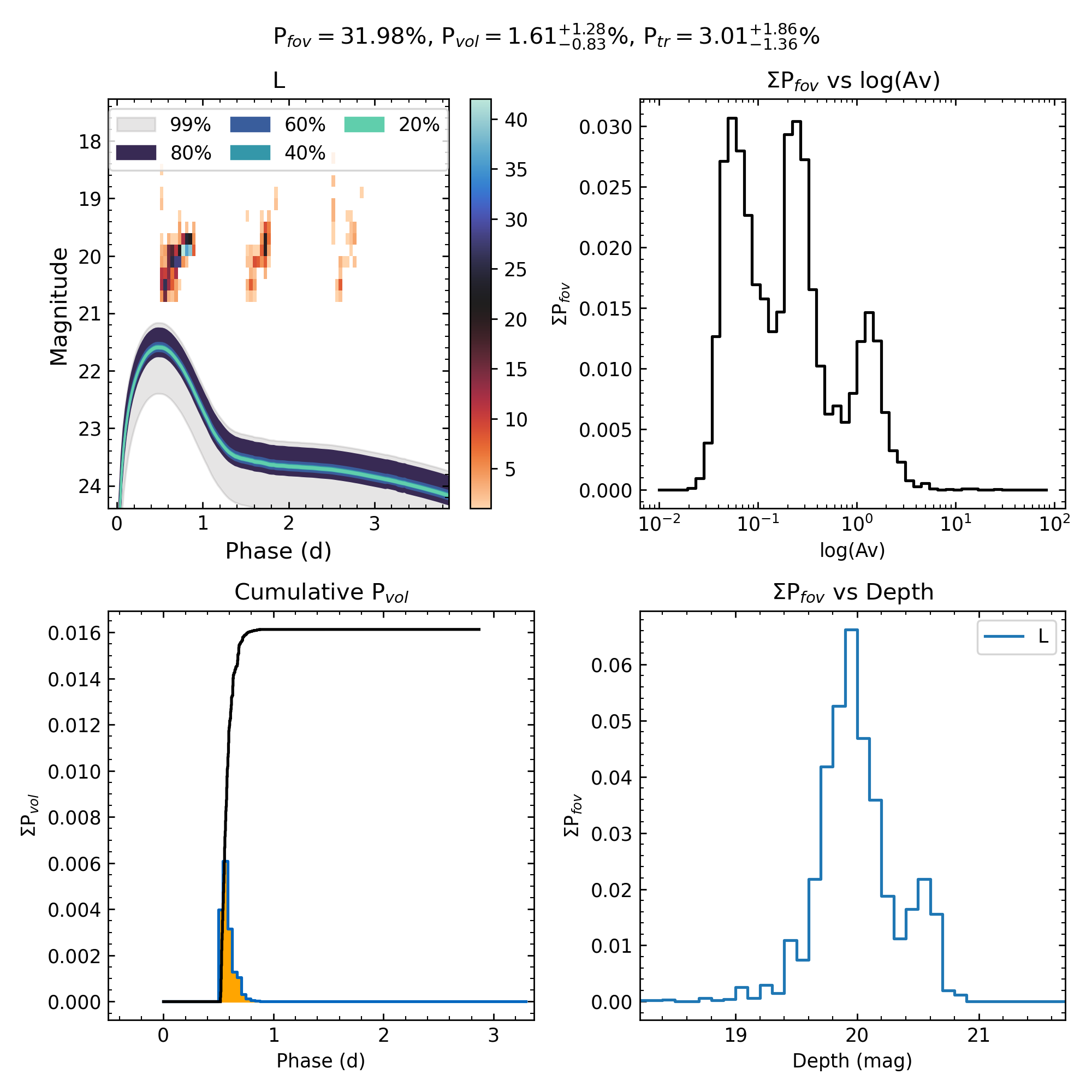}
    \caption{Plot showing the output from \textsc{GERry} detailing the results of the codes analysis of the follow-up of S190425z. \textbf{Top left:} Plot showing the expected light curve distribution and evolution, along with 2D histogram data showing the density of images binned into observation times and depths achieved. The light curve distribution is generated by placing the model light curves in all imaged skymap tiles and calculating the apparent magnitude based on the distance estimate contained within each skymap tiles. \textbf{Top right:} How the imaged 2D sky probability was distributed across milky way reddening. Most of the skymap suffered little extinction (A$_V<0.2$ mag). However some images covered the galactic plane resulting in very high extinction values. \textbf{Bottom left:} The probability volume (\pvol) evolution over the course of the campaign. In this case, the overwhelming majority of the recovered volume was done so in the first night of the campaign while a KN would've been at peak. The fading of the KN in the subsequent nights results in little to no new volume recovered. \textbf{Bottom right:} The imaged 2D sky probability distribution over image depth.}
    \label{fig:S190425z_result_plots}
\end{figure}

Figure \ref{fig:S190425z_result_plots} shows the expected light curve evolution along with the image depths achieved over the course of the campaign. While images were taken at the expected KN peak, they did not probe the required depth to recover the expected counterpart. This is primarily due to the distance of the source as the majority of the imaged probability suffered little dust reddening. The vast majority of the \pvol recovered on the first night when the KN was at peak, with the subsequent rapid fading rapidly diminishing the recovered probability on subsequent nights.

\begin{figure}
\centering
\includegraphics[width=0.6\columnwidth]{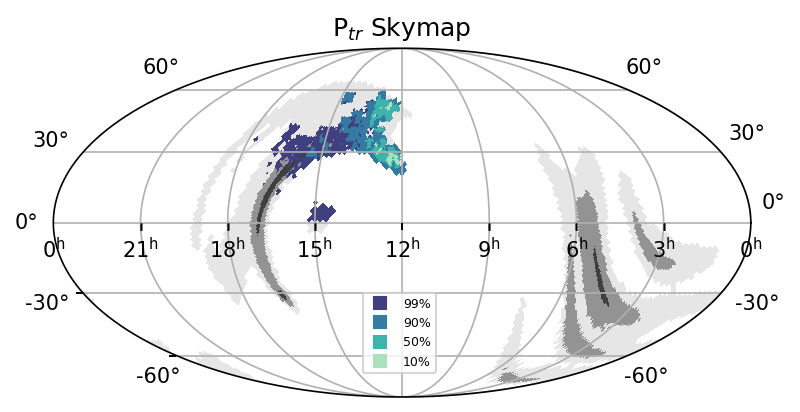}
    \caption{The \ptr skymap for the GOTO4 follow-up campaign of S190425z. While the follow-up covered 2155 \sqdeg, the 90\% credible area for a counterpart is $\sim$720 \sqdeg.}
    \label{fig:S190425z_ptr}
\end{figure}

Of the \ptr recovered (3\%), much of it was localised in the northern region of the sky where higher image depths were achieved close to the expected KN peak (figure \ref{fig:S190425z_ptr}. The viable area of the candidate is reduced from 2155 \sqdeg to 720 \sqdeg, and candidates can be ranked based on this statistic when presented to humans for verification, which would facilitate faster identification.

\section{Optimisation of follow-up strategy}

LIGO's current O4 observing period is define by two periods. The first, 'O4a' started on the 24th May 2023 and ended on the 16th January 2024. The first period consisted primarily of the LIGO Livingston (L1) and Hanford (H1) detectors both operating with a BNS sensitivity range $\sim$120-160 Mpc with Kagra (K1) contributing for a period of 1 month with a sensitivity of $\sim$1 Mpc. The second observing sub-period, 'O4b' started on the 10th April 2024 and is currently underway as of the time of writing with the planned end date of February 2025 and consists of both the previous aforementioned detectors along the Virgo (V1) which is operating with a sensitivity of 55-60 Mpc.

\begin{figure}
\includegraphics[width=\columnwidth]{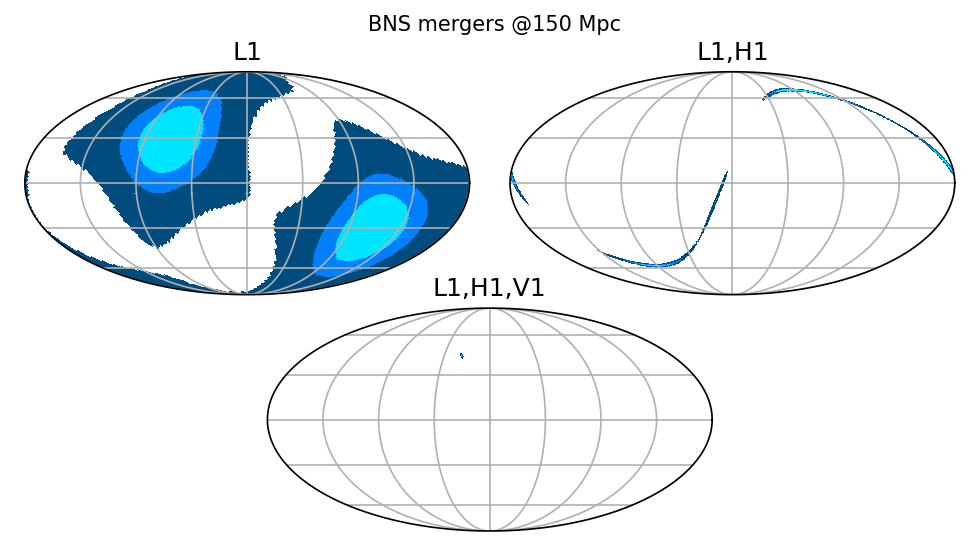}
    \caption{Plot showing the 90\%, 50\% and 25\% localisation areas on sky for different simulated BNS mergers at $\sim$150~Mpc using a single detector (L1), both LIGO detectors (L1 and H1) and the LIGO detectors with Virgo (L1, H1 and V1). These were generated using the methods outlined in section \ref{sec:simulations}.}
    \label{fig:detectors}
\end{figure}

The typical localisation area of a given gravitational wave relies heavily on triangulation of signals, and so depends strongly on the number of detectors online at any given time (see \ref{fig:detectors}). 
Events detected by a single detector are typically poorly constrained and localisation maps consist of a dipole with the localisation area typically comparable to half the sky ($\sim$21,000 \sqdeg). Two detector events are the most dynamic with localisation's ranging from small 'spots' on the order of 100s to 1000s \sqdeg to larger ring-like or 'banana' shaped profiles stretching across both hemispheres with localisation areas of 1000s to 10,000s \sqdeg depending on the SNR of the GW.
Which strategy (or strategies) to employ is a critical consideration if trying to detect any EM counterpart to these events as the effectiveness of a strategy could change considerably depending on the localisation skymap and the estimated distance to the event. 

Given that the primary objective of such follow-up is to detect EM counterparts, \ptr is an effective, quantitative measure of a given strategies performance assuming a model light curve. In this section we will demonstrate how using \textsc{GERry} in tandem with both simulated GW events and follow-up campaigns allows for an effective comparison of the effectiveness of observing strategies. We will also explore how a typical wide-field sky-survey like GOTO compares with the upcoming Vera C. Rubin Observatory (VRO).

\subsection{Simulation setup}
\label{sec:simulations}
The GW events were simulated using the \textsc{ligo-skymap}\footnote{https://pypi.org/project/ligo.skymap/} python package. One, two and three detector events were simulated up to distances of 300 Mpc in order to explore as much of the localisation area and distance parameter space as possible. The basic recipe for producing these events follow the instructions presented in the \textsc{ligo-skymap} documentation\footnote{https://lscsoft.docs.ligo.org/ligo.skymap/quickstart/bayestar-injections.html}, with only minor changes.
The simulations were set to trigger over a 1 year period from January 2024 over a uniform distance distribution up to 300 Mpc. The uniform distance (rather than volume) distribution was chosen so that closer distances ($\leq$ 100 Mpc) were being adequately sampled. In order to facilitate sampling over a wider range of localisation areas, the net GW SNR required for a `trigger' from the detectors was lowered from the default value of 12.0 to 5.0. We limit the neutron star mass range to 1.1--2.0 $M_{\odot}$ in line with estimates from P.~Landry \& S.~Jocelyn~(2021)\cite{NSmass}.
For each detector combination, a detection by all of the detectors is required. GW detector sensitivities were not scaled and so use the default values. One detector events only use L1, two detector events use L1 and H1 and finally three detector events use L1, H1 and V1; $\sim$73 one-detector, $\sim$270 two-detector, and $\sim$103 three-detector events were simulated.

The follow-up simulations were performed using a modified version \textsc{goto-tile} (M. Dyer\footnote{https://github.com/GOTO-OBS/goto-tile}), a package used by GOTO to define the follow-up schedule of a skymap. \textsc{goto-tile} takes basic telescope information such as field-of-view, location, number of mounts and an pointing/exposure time to simulate the follow-up of a probabilistic skymap over a set duration. A fixed grid of tiles is generated on sky according to the field of view of the telescope. When used with a skymap, the code identifies which tiles cover the associated 95\% probability contour. At a time step corresponding to the typical pointing time, tile visibility is calculated based on twilight and altitude constraints. From the remaining visible tiles, those which have been imaged least are prioritised. For tiles with the same number of observations, the contained probability and airmass are calculated and a tiebreaker is calculated by weighting the airmass using the contained probability.

GOTO is now at design specification, with four mounts each having a field of view of $\sim$44\,deg$^2$: two mounts at Roque de los Muchachos Observatory in La Palma and a further two at Siding Spring Observatory in Australia. All simualated observations were taken in the wide L-band filter. We explore two strategies, using shallow $4\times90$\,s exposure sets per pointing to maximise 2D coverage and a deeper $12\times60$\,s exposure set to maximise recovered volume per pointing. A 60\,s overhead is assumed at each pointing to simulate slewing, rapid focusing and readout procedures giving a total of 420\,s and 780\,s for each strategy respectively. The image depths at each pointing were informed by archival GOTO image data. Images from the GOTO database for each of the above exposure times were queried at different airmasses. A curve was fit to airmass and average depth values which was then used to estimate the appropriate depth for each simulated image. In the case of the deeper $12\times60$\,s strategy this results in a depth of 20.6 and 20.26\,mag at airmass = 1 and 2, respectively. The shallower $4\times90$\,s strategy had depths of 20.12 and 19.45\,mag at corresponding airmasses.

For VRO, we used a single mount with 9.6 \sqdeg field of view located at Cerro Pachón in Chile. Images were simulated in the 5 loaded filters (\textit{g,r,i,z,y}), each with 30\,s exposures. While VRO images are somewhat circular or hexagonal in nature, we simply assume a square field of view with the same total field of view area. We use the overheads detailed in Jones, R.~L. et al.~(2020)\footnote{Survey Strategy and Cadence Choices for the Vera C. Rubin Observatory Legacy
Survey of Space and Time (LSST)}. Along with four filter change times of 120s each with an additional 3 seconds for each readout and 30s for slew time for each pointing results in a total pointing time of 675s. We also simulate a three day follow-up duration, although this is unlikely in reality as targeted GW follow-up is not a primary science goal of VRO. However, it allows for the assessment of what is theoretically achievable from an 8~m class survey telescope. The strategy for VRO follow-up of GW events is not yet finalised, and so pointing strategy we use here is representative (an alternative could save filter change time at the expense of less colour information, for example). Given VRO has not yet achieved first light, we have no real data to sufficiently model the depth/airmass relation therefore we simply use the values for single exposures in each filter detailed here\footnote{https://www.lsst.org/scientists/keynumbers}. This is likely going to result in an over-estimate as it neglects the effects of sub-optimal observing conditions, therefore we will leave much more detailed modelling to other studies. 

In order to estimate light curve properties and evolution for a typical KNe we again used \textsc{MOSFiT}\cite{mosfit}. We used the listed input priors found in table 1 of M. Nicholl et al.~(2021)\cite{mosfit170817}, with the exception of limiting chirp mass $M$ and mass ratio $q$ to correspond to the aforementioned neutron star mass range of 1.1--2.0 $M_{\odot}$. 10,000 light curves were generated and and the average absolute magnitude light curve was calculated. For all of the following simulations, unless otherwise stated, we disabled the effects of reddening as we primarily want to assess the effects of localisation area and distance on the instruments and strategies rather than Galactic effects.

\begin{figure}
 \centering
\raisebox{1.2cm}{\includegraphics[width=0.59\columnwidth]{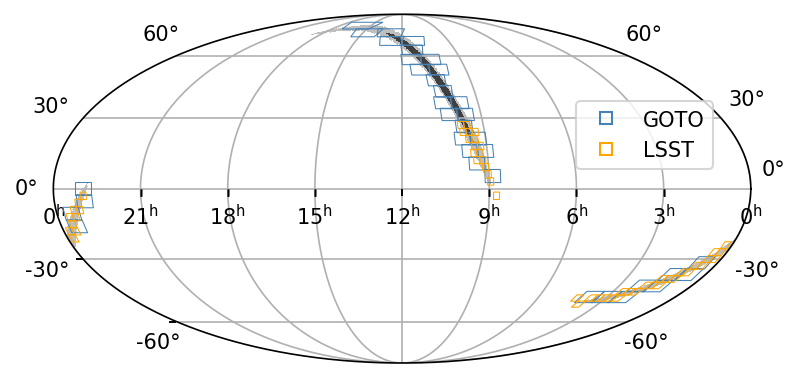}}
\includegraphics[width=0.4\columnwidth]{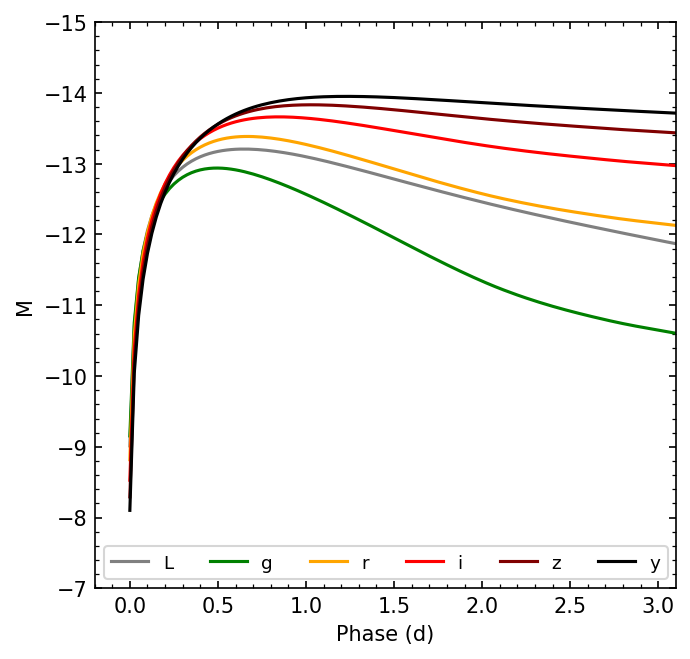}
    \caption{\textbf{Left:} Simulated localisation skymap and observational follow-up campaign of a two-detector GW event using GOTO (blue) and VRO (orange). \textbf{Right:} The fiducial KN lightcurve in GOTO L-band as well as the filters used in the VRO simulations.}
    \label{fig:sims}
\end{figure}

An example of a simulated follow-up campaign and the light curves used can be seen in figure \ref{fig:sims}. The light curves peak at $M = -13$ in L and $g$, almost two magnitudes fainter than KN 2017gfo. How representative KN\,2017gfo is of the KN population is poorly constrained, although it appears well withing the range of luminosities seen for other candidate KNe, if somewhat on the luminous side\cite{BenDiversity}. There may also be shortcomings of the simulated light curves, but we are primarily using them for our strategy-informing analysis where the observing campaign is the variable factor.

\subsection{Results: GOTO strategy}

\begin{figure}
\centering
\includegraphics[width=0.75\columnwidth]{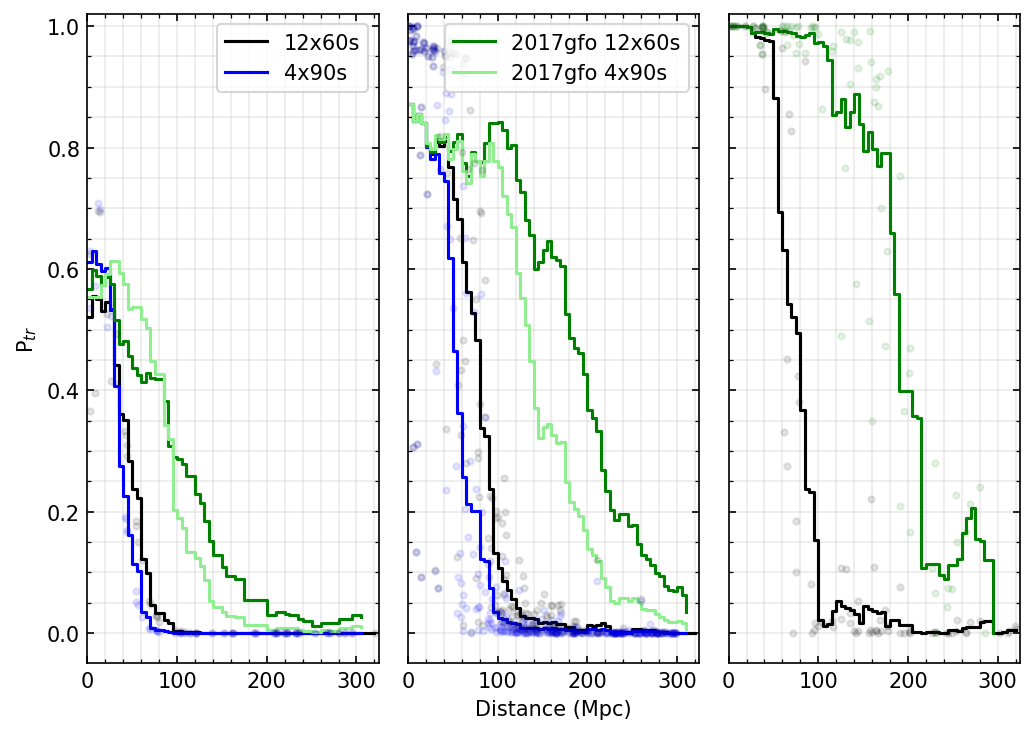}
    \caption{GOTO \ptr recovery over distance for (left to right) one, two and three detector scenarios. Shown here is a rolling average (lines) of individual simulations (points) using 25~Mpc bins in 5~Mpc intervals to minimise fluctuations due to distance bin choices. Black lines show the results of the deeper $12\times60$\,s strategy and blue the shallower, but faster $4\times90$\,s strategy.  As a comparison, we also show the affect of using a model light curve of KN 2017gfo instead of the `average' KN light curve. For 3 detector events we do not simulate the faster strategy as the very small localisation areas (typically 0.1--10\sqdeg) are already covered by a single GOTO pointing and so, offers no advantages. The unusual peak at 280~Mpc is the result of skymaps containing pixels with closer distance estimates that deviate from the typical distance distribution, resulting in higher \ptr values than expected at larger distances.}
    \label{fig:GOTO_results}
\end{figure}

The results of the simulations and the \textsc{GERry} analysis are shown in Figure \ref{fig:GOTO_results}. Rather unsurprisingly, increasing the number of detectors results on average higher \ptr values as it becomes easier to completely cover the skymap. For the one detector scenarios, the faster $4\times90$\,s strategy outperforms the $12\times60$\,s strategy for BNS events less than 30~Mpc assuming an `average' KN light curve. However, the deeper strategy starts to rapidly outperform the shallower one at larger distances as the primary limiting factor changes from 2D sky coverage to depth. For AT~2017gfo-like light curves, this threshold is increased to $\sim$100~Mpc

\begin{figure}
\centering
\includegraphics[width=0.49\columnwidth]{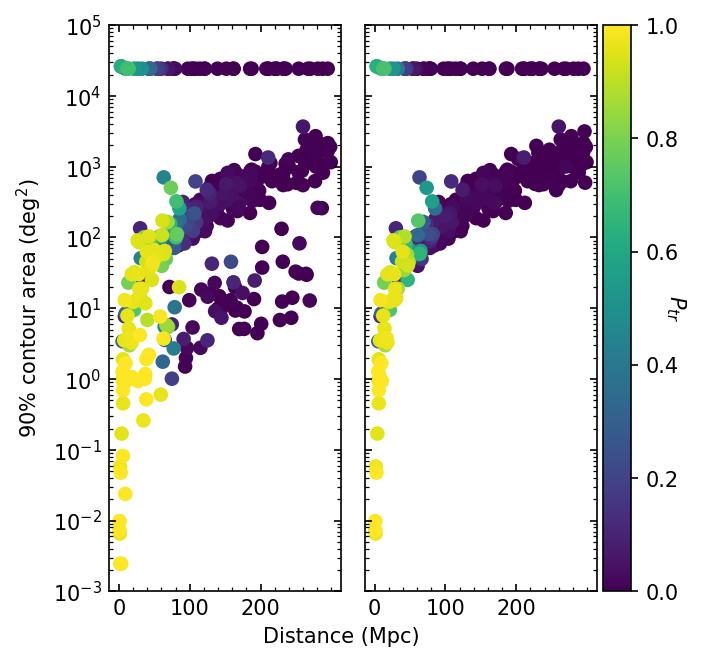}
\includegraphics[width=0.49\columnwidth]{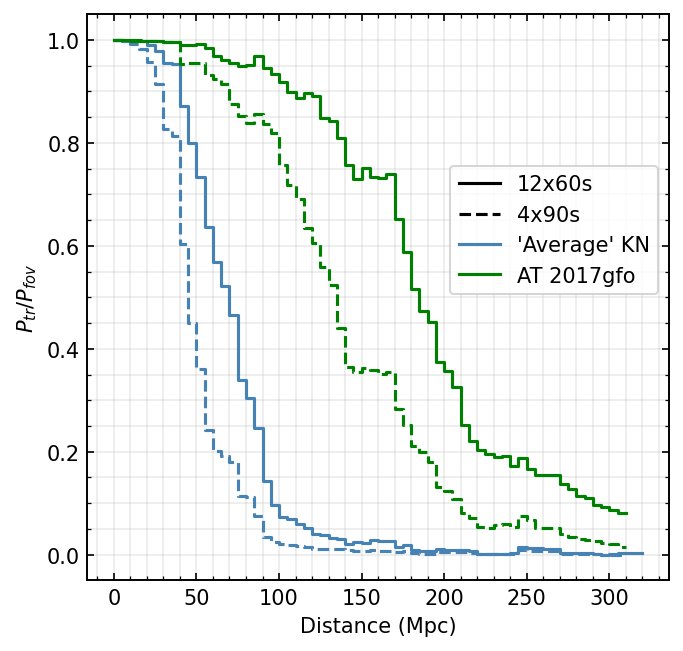}
    \caption{\textbf{Left:} Plot showing the size of the 90\% credible area vs distance with \ptr mapped to the point colours for the deeper $12\times60$\,s strategy and the shallower $4\times90$\,s strategy. \textbf{Right:}Normalised \ptr values for each of the strategies showing only the effects of distance on \ptr recovery.}
    \label{fig:CvsD}
\end{figure}

For two and three detector events, the size of the 90\% credible area only has a minor effect as shown in Figure \ref{fig:CvsD}. GOTO's field of view is large enough that it is able to fully image skymaps with localisation areas in the order of 10--100s \sqdeg with relative ease. Similarly to the one detector scenario, for the 'average' KN light curve the deeper strategy is preferable for GW events at $\ge$30Mpc. If KNe are more similar to AT~2017gfo we can see that both the shallower and deeper strategies perform similarly up to $\sim$80 Mpc. We do not test the shallower strategy on three-detector events as their localisation is significantly less than the field of view of a single GOTO pointing, therefore it will not offer any increase in cadence or skymap coverage.
In the three detector scenario, an unusual peak is visible at 280~Mpc. Typically GW skymap pixels have gaussian-like distance distributions. The simulated skymaps at times can display more irregular or varied distributions such as bimodal distributions. The \ptr peak is caused by skymaps with a strong 'closer' distance peak that results in higher \ptr values being recovered for distant skymaps. This is much less prominent in the 'average' light curve results as it is much fainter and is unlikely to be recovered even with the closer distances.
From this analysis we can conclude that the shallower strategy is preferable for events $\le$30~Mpc as it offers superior cadence for similar performance values while the deeper strategy should be employed for more distant events. 
To isolate the effects of GW distance/depth for each strategy, we can combine the results for all detector combinations for both strategies and normalise the average \ptr value using \pfov, which is shown in the right panel of Figure \ref{fig:CvsD}. Here we can more clearly see the true effective distance horizons for each strategy.

Given that \ptr represents the probability of detecting a KNe, we can convolve the curves in Figures \ref{fig:GOTO_results} with the expected volumetric rate to obtain estimates for the total number of detected KNe over a given period. 

\begin{figure}
\centering
\includegraphics[width=0.6\columnwidth]{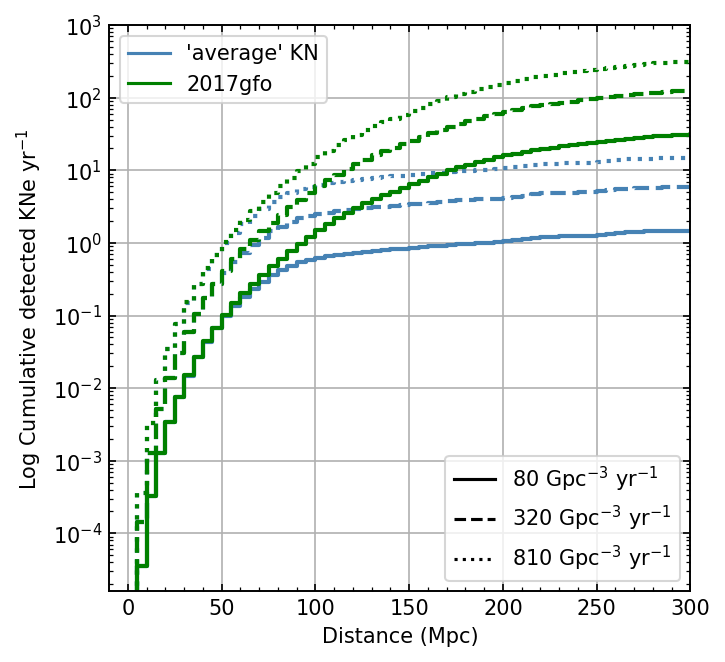}
    \caption{The estimated rates of detected KNe for GOTO assuming a AT~2017gfo-like light curve as well as the more `pessimistic' average KN light curve using the 12x60s strategy.}
    \label{fig:Rates}
\end{figure}

Assuming the following: (1) localisations from the two-detector scenario ($<$1000\,\sqdeg out to 300\,Mpc), (2) the BNS merger rate is 80--810 Gpc$^3$yr$^{-1}$ as reported by R. Abbott (2021e) \cite{BNSrates}, (3) the more pessimistic 'average' KN scenario is representative of the KN population, and (4) there is no downtime for both GW interferometers and GOTO, we estimate that GOTO could recover 1 to 11 KNe per year using the 12x60s strategy. If KNe are more similar to AT~2017gfo, the recovery rate increases by a factor of 10. Given the significant uncertainty in the inferred BNS merger rates, we plot the estimates and bounds provided in in Figure \ref{fig:Rates}. This should be however, treated as a hard upper limit as both GW detectors as well as telescopes are taken offline regularly for maintenance. The GW detector BNS horizon also fluctuates depending on number and combination of GW detectors that are online. Telescopes also have to contend with observing time losses due to weather.

\subsection{Results: VRO}

\begin{figure}
\centering
\includegraphics[width=0.6\columnwidth]{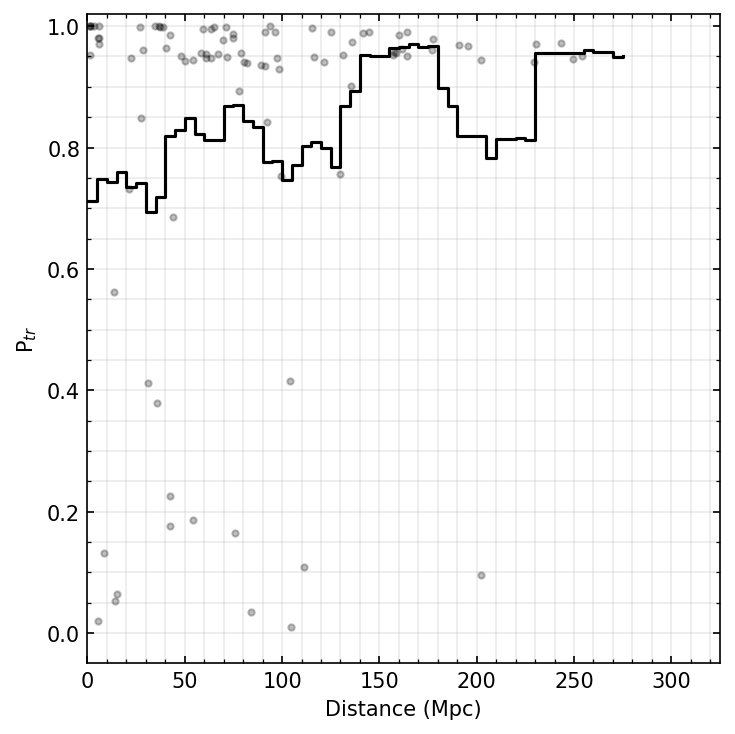}
    \caption{VRO follow-up \ptr recovery for a three-day follow-up versus source distance for two- and three-detector GW events with localisations $\leq$150\,\sqdeg. The pessimistic `average' KN light curve model has been used.}
    \label{fig:LSST_results}
\end{figure}

The results for VRO are shown in Figure \ref{fig:LSST_results}. It is unlikely VRO will trigger on GW events with large sky localisations ($\gtrsim$100s\,\sqdeg) due to the comparatively modest field of view and it not being core science of the observatory. We therefore limit the simulated events to two- and three-detector skymaps with localisation areas $\leq$150\,\sqdeg. The results show significant scatter in comparison to GOTO. This is almost solely driven by the skymap visibility from the VR, and we note here we are only showing results for events that have at least some part of their 95\% localisation areas visible to VRO.
Overall the results show that suffers little to no drop-off as a function of distance as VRO is able to reach the image sensitivity required for distant GW events, even for this faint `average' KN.

\begin{figure}
\centering
\includegraphics[width=0.6\columnwidth]{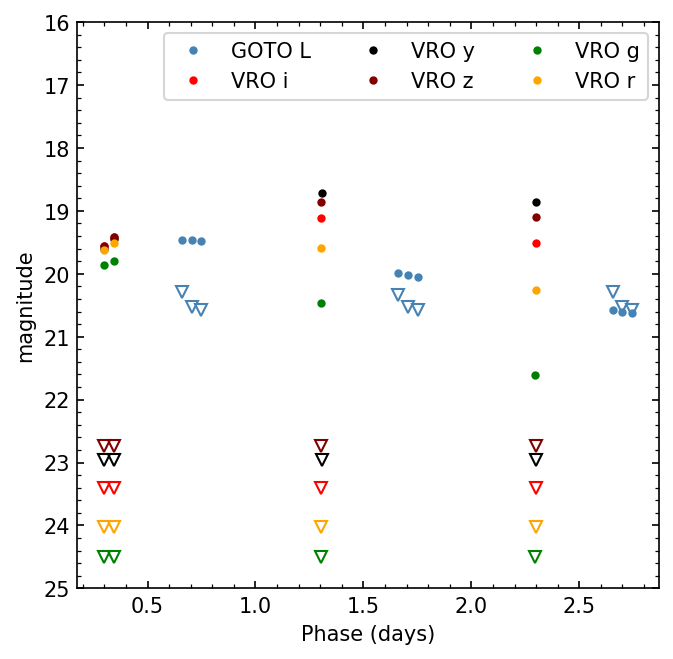}
    \caption{An example of the expected `observed' light curves from the GW and follow-up simulations of a two detector GW event (446) at 63~Mpc assuming an 'average' KN. Also shown are the simulated image depths (open triangles).}
    \label{fig:gerry_lightcurve}
\end{figure}

It is possible to examine what the 'observed' light curve of the counterpart would have looked like during a given follow-up campaign at any imaged sky position. Using the \textit{"processing\_dfs = True"} flag, we can use the output \textit{imout} csv file which contains the UNIQ skymap tile ids and the counterpart magnitudes contained within. By converting an RA and DEC to UNIQ tile id (see here\footnote{https://rtd.igwn.org/projects/userguide/en/latest/tutorial/multiorder\_skymaps.html}). one can filter the csv file and plot the respective 'phase' and 'mag' columns. An example is shown in figure \ref{fig:gerry_lightcurve} for both GOTO and VRO follow-up of the same skymap, at the same sky position.

\subsection{VRO strategy}
As mentioned VRO will likely perform more modest GW follow-up, involving a limited amount of time spent on a sub-set of opportune events, rather than continuously observing for nights uninterrupted post-trigger. We examine one of the strategies outlined in I. Andreoni et al.~(2022)\cite{VROstrat}, specifically the `minimal strategy': one coverage of the 90\% localisation area with 30s exposures in $g + i$, followed by a second night of 180s exposures. Here we assume depths m$_{g}=26.0$ and m$_{i}=25.0$ using their figures 1 and 2. This strategy will be employed for skymaps with localisation areas between 20--100\sqdeg.
We select skymap simulation 301 with a localisation area of 59~\sqdeg at a distance of 64~Mpc. The `unrestrained' campaign resulted in equivalent \pfov and \ptr values of 94.7\%, since almost the entirety of the visible probability volume was recovered (under which conditions \pfov = \ptr). We find that the minimal strategy outlined above returns almost identical \pfov, \pvol and \ptr values demonstrating it is more than capable of detecting a KN  considerably more conservative survey strategy. As noted by I Andreoni et. al, the expected time investment using their \textit{preferred} strategy for 20--100\sqdeg localisations is 5--6 hours. Using \textsc{GERry} we find that promising \ptr values ($\ge$90\%) can be achieved for a skymap with localisation area ~187\sqdeg at a distance of $\sim$143~Mpc by using the \textit{minimal} strategy above, even reduced to 30s exposures across both nights totalling 5.4 hours, potentially increasing the number of trigger-able events. While a single night of 30s is capable of producing high \ptr values, having multiple epochs of detections to map rise/decline rates is important in separating a counterpart such as a KN from other, more common transients such as supernovae.


\subsection{Combining community efforts in follow-up}
\begin{figure}
\centering
\includegraphics[width=0.6\columnwidth]{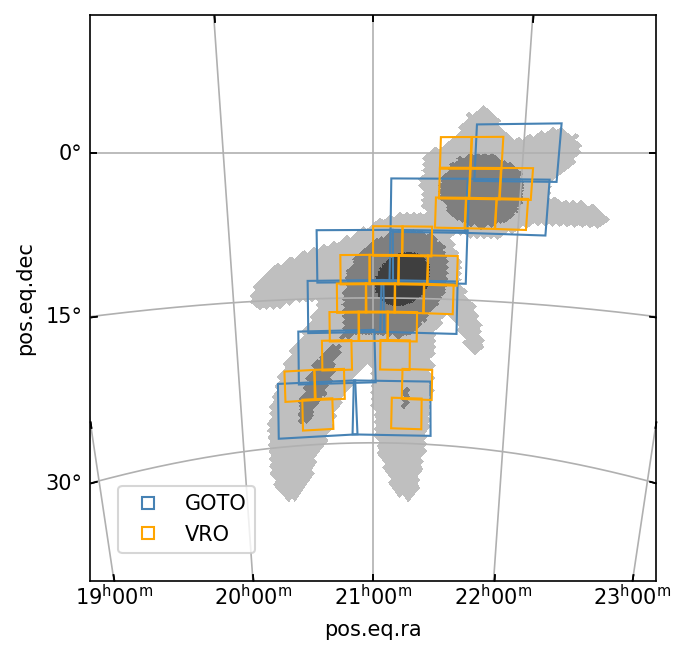}
    \caption{The footprints of the simulated GOTO and VRO observations of simulation 245.}
    \label{fig:combined_foot}
\end{figure}

\textsc{GERry} is agnostic to the origin of the images it analyses. It can effectively and correctly combine image data from multiple follow-up campaigns into a single bulk analysis. As an example, shown in figure \ref{fig:combined_foot} is the follow-up of simulation 245, a BNS merger at 69~Mpc. Individually GOTO and VRO achieved \ptr values of 86.14\% and 93.5\% respectively. Combined, this value increased to 97\%. 
GOTO and VRO are only two of an increasing number of wide-field survey instruments triggering on GW events. By folding in not only the efforts of GOTO and VRO, but also of ATLAS, BlackGEM, DECam, Pan-STARRS, ZTF, and others, \textsc{GERry} can provide an quantitative and holistic description of the EM community's effort in follow-up up these rare event. The outputs can be used to perform a post-mortem and identify gaps in our combined capabilities and strategies. Such constraints, in the face of further non-detection follow-up efforts, will eventually allow for an assessment of EM model compatibility, helping to further our understanding of the nature and diversity of these events.

\section{Conclusion}
Multi-messenger astronomy is still a relatively new and rapidly growing frontier in time domain astronomy. In the case of finding the EM counterparts to BNS mergers, many challenges remain despite increasing number and sophistication of our instrumentation. In this paper we have shown how \textsc{GERry} can be used to help observers overcome some of these challenges. Starting from the initial steps of developing an observing strategy, we have shown how, through the use of GW and follow-up simulations, one can optimise observing strategy in key parameter spaces. We have also shown how the code can be used on real GW events to assess the follow-up campaign's performance, identify whether a counterpart was likely to be detected, as well as narrowing down the possible sky positions of the detected counterpart.
\textsc{GERry} is a simple but effective high-level product that can play a key part in informing both collaborations internally and the community at large about past or ongoing follow-up efforts. Particularly, it can inform on the effectiveness of our combined strategies if integrated in tandem with other public tools such as the GW Treasure Map\cite{treasuremap}. By analysing the combined efforts of many surveys using \textsc{GERry}, one can begin placing tighter constraints on counterpart emission than is possible from each individual survey's follow-up campaign, from which future models could be more securely anchored.
The development of the code continues with the aims of improving and streamlining the user experience.


\section{Availability}
All of the \textsc{GERry} results files and simulation data used to make the plots will be hosted on the Github.
The public release of \textsc{GERry} itself is planned via Github, and is available via reasonable request to the corresponding author.

\subsection* {Acknowledgments}
DO and JL acknowledge support from a UK Research and Innovation Fellowship(MR/T020784/1).
The Gravitational-wave Optical Transient Observer (GOTO) project acknowledges the support of the Monash– Warwick Alliance; the University of Warwick; Monash University; the University of Sheffield; the University of Leicester; Armagh Observatory \& Planetarium; the National Astronomical Research Institute of Thailand (NARIT); the Instituto de Astrof \'isica de Canarias (IAC); the University of Portsmouth; the University of Turku; the University of Manchester and the UK Science and Technology Facilities Council (grant numbers ST/T007184/1, ST/T003103/1, and ST/T000406/1).
DO acknowledges The Seventh Workshop on Robotic Autonomous Observatories as part of the Revista Mexicana de Astronomia y Astrofisica Conference Series (2024), which a preliminary version of this work will appear.


\bibliography{report}   
\bibliographystyle{spiejour}   





\end{spacing}
\end{document}